\documentclass[12pt,a4paper]{article}
\usepackage{graphics}
\usepackage{epsfig}
\usepackage{latexsym}
\usepackage{psfrag}
\usepackage{amssymb, amscd}

\newcommand{\Det}{\mathop{\rm Det}\nolimits}
\newcommand{\tr}{\mathop{\rm tr}\nolimits}

\begin{document}
 
\title{\begin{flushright}
{\small Preprint HU--EP--07/02 }
\end{flushright}
\bf Color superconductivity in the static Einstein
  Universe}
 
\date{ }
\author{D.~Ebert$^{1}$\footnote{debert@physik.hu-berlin.de},
  A.~V.~Tyukov$^{2}$, and
  V.~Ch.~Zhukovsky$^{2}$\footnote{zhukovsk@phys.msu.ru}\\
\small\sl $^{1}$ Institut f\"ur Physik, Humboldt-Universit\"at zu Berlin, 12489 Berlin, Germany\\
\small\sl $^{2}$ Faculty of Physics, Department of Theoretical Physics, \\
\small\sl Moscow State University, 119899, Moscow, Russia
}
\date{\today}
\setcounter{page}{1}
\maketitle
 
\begin{abstract}
We study the behavior of quark and diquark condensates in dense quark
matter under the influence of a gravitational field
adopting as a simple 
model the static
$D-$dimensional Einstein Universe. 
Calculations are performed in the framework of
the extended Nambu--Jona-Lasinio model  at finite temperature
and quark density on the basis of 
the thermodynamic potential and the gap equations.
Quark and diquark 
condensates as functions of the chemical potential
and temperature at different values of the curvature  have been
studied. Phase portraits of the system have been constructed.
\end{abstract}
 
 
\maketitle

\section{Introduction}
 
As is well known, non-perturbative effects in low energy QCD can be
studied with the use of various effective models with
four-fermion interaction, such as the instanton model \cite{rapp} or
the Nambu--Jona-Lasinio model (NJL)~\cite{NJL}.
 
Effective field theories with four-fermion
interaction of the  Nambu -- Jona-Lasinio type, which
incorporate the phenomenon of dynamical chiral symmetry breaking, are
quite useful in describing  the physics of light mesons (see
e.g. \cite{MESONS,generalN,kunihiro} and references therein) and
diquarks \cite{Ebert_Kaschluhn, Vogl}. Moreover, on their basis one
can
study the effects of various external conditions, like
temperature and chemical potential~\cite{Kawati}, and
consider the influence of external electromagnetic and gravitational
fields.  In
particular, the role of the NJL approach increases,
when detailed numerical lattice calculations are not yet admissible in
QCD with nonzero baryon density and in the
presence of external gauge fields
\cite{kl,vvk,ekvv,ek,Ebert_Zhukovsky}. Moreover, the NJL model has also
applications in the nuclear physical  and
astrophysical researches (e.g., neutron stars) as concerns the quark
matter~\cite{quark_matter}.
 
It was proposed more than twenty years
ago \cite{Ba,Frau,bl} that at high baryon densities a colored
diquark condensate $<qq>$ might appear.  
In analogy with ordinary
superconductivity, this effect was called color superconductivity
(CSC).  In particular, the CSC phenomenon was investigated in the framework of the one-gluon
exchange approximation in QCD~\cite{one_gluon}, where
the formation of diquarks 
(colored Cooper pairs of quarks) 
is predicted self-consistently
at extremely high values of the chemical potential $\mu\gtrsim 10^8$
MeV~\cite{extream}. One should also mention the use of the
Bethe-Salpeter equation to obtain the diquark masses in the CSC phase
of cold dense QCD at asymptotically large values of the chemical
potential~\cite{26}. Unfortunately, the corresponding high baryon densities are not
observable in nature and not accessible in experiments. In order to study the
problem at lower values of $\mu$, various effective
theories for low energy QCD, such as the instanton model  \cite{rapp}
and the Nambu$-$Jona-Lasinio (NJL) model\cite{NJL} can be employed.
The
possibility for the existence of the CSC phase in the region
of moderate densities was recently proved
(see, e.g., the papers~\cite{rapp,Alford,Berges,Schwarz},
as well as the review articles~\cite{Alford_Kebrikov,shovk}
and references therein).
In these papers it was shown
that the diquark condensate $<qq>$ can appear  already at a rather
moderate 
baryon density ($\mu\sim 400$ MeV). The conditions favorable for this
condensate to be formed can possibly  exist in the cores of cold
neutron stars. 
Since quark Cooper pairing occurs in the color 
anti-triplet
channel, a nonzero value of $<qq>$ means that, apart from
the electromagnetic $U(1)$ symmetry,  the color
$SU_c(3)$ symmetry should be spontaneously broken  inside
the CSC phase as well. In the framework of NJL models the CSC phase formation has generally
been considered
as a dynamical competition between diquark  $<qq>$ and
usual quark-antiquark condensation $<\bar qq>$. Special attention has
been paid to the catalyzing influence of the vacuum color fields on
the condensation of diquarks \cite{Ebert_Zhukovsky} (see also,
\cite{braghin}).
Moreover, mesons and diquarks in
the color neutral superconducting phase of dense cold quark matter
were studied in \cite{yud} and
Nambu-Goldstone bosons in the color-asymmetric
superconducting phase were considered in the framework of the
Nambu--Jona-Lasinio-type model in \cite{blasch}.
 
In several papers, in the framework of the NJL model, the influence of
a gravitational field on the dynamical chiral symmetry breaking
(D$\chi$SB) due to the
creation of a finite quark condensate $<\bar{q}q>$ has been
investigated at zero values of temperature and chemical
potential~\cite{MUTA,
Elizalde,Elizalde_Shilnov,Gorbar}. The study of the combined influences
of curvature and temperature has been performed
in~\cite{Inagaki_Ishikawa}.  Recently,  the
dynamical chiral symmetry breaking and its restoration
for a uniformly accelerated observer  due to the thermalization effect  
of acceleration was studied in~\cite{ohsaku2} at zero chemical
potential.  Further investigations of
the influence
of the Unruh temperature on the phase transitions in dense quark
matter with a finite chemical potential, and especially on the restoration
of the broken color symmetry in CSC were made in \cite{qqrindler}.
One of the widely used methods of accounting for the gravitation is
based on the
expansion of the fermion propagator in powers of small
curvature~\cite{Bunch_Parker,Parker_Toms}. 
For instance, in \cite{kim_klim}, the 
three-dimensional Gross-Neveu model in a space-time 
with a weakly curved two-dimensional surface was investigated, using
an effective potential at finite curvature and nonzero chemical
potential. In 
paper~\cite{Goyal_Dahiya}, this approximation was used in considering
the D$\chi$SB at non-vanishing temperature and chemical potential.
It should, however, be mentioned that near the
phase transition point, one cannot consider 
the critical curvature
$R_c$ to be small and
therefore exact
solutions with finite values of $R_c$ should be used. This kind of
solution with consideration for the chemical potential and temperature
in the gravitational background of a static Einstein universe has
recently been considered
in~\cite{Huang_Hao_Zhuang}. In all these papers it was
demonstrated that chiral symmetry is restored at high values of the
space curvature. 
There arises the interesting question, whether
similar effects can occur in the study of
CSC under the influence of a gravitational field.
 
In the present paper, we study the influence of temperature, chemical
potential and a gravitational field on the creation of the CSC state in
quark matter. As is well known, one of the possible models for
the expanding universe is the closed Friedman
model. However, since the formation of condensates 
is expected to take place
considerably faster than the rate of the Universe expansion, its
radius in our calculations can be considered as constant. Therefore,
as a simple
model of gravitation with a given background space-time
we assume a static Einstein universe.
In this case, we will derive an analytical expression for
the effective potential of the model
by using the mean field
approximation, with all the information concerning the chiral and
color condensates. As is known, in a flat 
space-time, the chemical
potential and temperature lead to fairly different effects, i.e., at
large chemical potential, a first order phase transition takes
place, while high temperatures lead to the second order phase
transition. The aim of this paper is to study the phase structure of
the quark matter, in particular CSC, in a highly curved 
space-time 
in the model of the static Einstein universe.
 
\section{The extended NJL model in curved space-time}
 
In  D-dimensional curved space-time with signature $(+,-,-,-,\cdots)$,
the line element is written as
\[
ds^2=\eta_{\hat a\hat b}e_\mu^{\hat a}e_\nu^{\hat b}dx^\mu dx^\nu.
\]
The gamma-matrices $\gamma_{\mu}$, metric $g_{\mu\nu}$ and the
vielbein $e^{\mu}_{\hat{a}}$, as well as the definitions of the 
spinor covariant derivative $\nabla_{\nu}$ and  
spin connection $\omega^{\hat{a}\hat{b}}_{\nu}$ are given by the following
relations~\cite{Parker_Toms,brill}:
\begin{eqnarray}
& & \{\gamma_{\mu}(x),\gamma_{\nu}(x)\}=2g_{\mu\nu}(x), \quad
  \{\gamma_{\hat{a}},\gamma_{\hat{b}}\}=2\eta_{\hat{a}\hat{b}}, \quad
  \eta_{\hat{a}\hat{b}}={\rm diag}(1,-1,-1,-1,\cdots), \nonumber \\
& & g_{\mu\nu}g^{\nu\rho}=\delta^{\rho}_{\mu}, \quad
  g^{\mu\nu}(x)=e^{\mu}_{\hat{a}}(x)e^{\nu \hat{a}}(x), \quad
  \gamma_{\mu}(x)=e^{\hat{a}}_{\mu}(x)\gamma_{\hat{a}}. \label{2}\\
 & &
  \nabla_{\mu}=\partial_{\mu}+\Gamma_{\mu},\quad\Gamma_{\mu}=\frac12\,\omega^{\hat a \hat b}_{\mu}\sigma_{\hat a \hat
  b},
  \quad
   \sigma_{\hat{a}\hat{b}}=\frac{{1}}{4}[\gamma_{\hat{a}},\gamma_{\hat{b}}],
 \nonumber \\ 
& & 
\omega^{\hat{a}\hat{b}}_{\mu}=\frac{1}{2}e^{\hat{a}\lambda}
e^{\hat{b}\rho}[C_{\lambda\rho\mu}-C_{\rho\lambda\mu}-C_{\mu\lambda\rho}],
 \quad C_{\lambda\rho\mu}=
e^{\hat{a}}_{\lambda}\partial_{[\rho}e_{\mu]\hat{a}}.
\label{3}
\end{eqnarray}
Here, the index $\hat{a}$ refers to the flat tangent space defined by the
vielbein at 
space-time 
point $x$, and the $\gamma^{\hat{a}}
(\hat a=0,1,2,3,\cdots)$ are the usual Dirac gamma-matrices  of Minkowski
space-time.
Moreover
$\gamma_5$,  is defined,  
as usual (see, e.g., \cite{Parker_Toms,bordag,Camporesi}), 
i.e. to be the same as in
flat space-time and thus independent of space-time variables.  
In the following we shall consider the static Einstein universe
with the line element
\begin{equation}
\label{metric}
  ds^2=dt^2-g_{ij}(\vec{x})dx^i dx^j \,\,\, (i,\,j=1,\ldots, D-1).
\end{equation}
 
A conventional model that demonstrates the formation of the color
superconducting phase is the extended NJL model with up- and
down-quarks. This model may be considered as the low energy limit of
QCD. For the color group $SU(3)_c$ its Lagrangian takes the form \footnote{In this work, we shall consider the
  particular realistic case of $N_c=3$. The general case of arbitrary
  $N_c$ is discussed in \cite{generalN} with relation to the
  effective hadron theory of QCD. In what follows, however, we shall
  keep the general notation $N_c$ for convenience.}
\begin{equation}
\begin{array}{ll}
    \label{lagrangian}
    \mathcal{L} &=
    \bar{q} \left[ i \gamma^{\mu}\nabla_{\mu} + \mu \gamma^{0} \right] q
    + \frac{G_{1}}{2N_{c}}
    \left[
        \left( \bar{q} q \right)^{2}
        + \left( \bar{q} i \gamma^{5} \vec{\tau} q \right)^{2}
    \right] +{}\\
    &+ \frac{G_{2}}{N_{c}}
    \left[
        i \bar{q}_{c} \varepsilon\epsilon^{b} \gamma^{5} q
    \right]
    \left[
        i \bar{q} \varepsilon\epsilon^{b} \gamma^{5} q_{c}
    \right].
\end{array}
\end{equation}
Here, $\mu$ is the quark chemical potential,
$q_c=\textit{C}\bar{q}^t,\bar{q}_c=q^t\textit{C}$\,\, are
charge-conjugated bispinors ($t$ stands for 
the transposition
operation). The charge conjugation operation is
  defined, as usual (see, e.g., \cite{Parker_Toms}), with the help of
  the operator  $\textit{C}=i\gamma^{\hat   2}\gamma^{\hat 0}$, where
  the flat-space matrices $\gamma^{\hat   2}$ and $\gamma^{\hat 0}$
  are used. 

The quark field
$q\equiv q_{i\alpha}$ is
a doublet of flavors and
triplet of colors with indices $i=1,2;\;\alpha=1,2,3.$ Moreover,
$\vec{\tau}\equiv(\tau^1,\tau^2,\tau^3)$ denote Pauli matrices in the
flavor space;
$(\varepsilon)^{ik}\equiv\varepsilon^{ik},\;(\epsilon^{b})^{\alpha\beta}\equiv\epsilon^{\alpha\beta
  b}$
are the totally antisymmetric tensors in the flavor and color spaces,
respectively.
 
Next, by applying the usual bosonization procedure, we obtain the
linearized version of the Lagrangian~(\ref{lagrangian}) with
collective boson fields $\sigma$, $\vec\pi$  and $\Delta$,
\begin{equation}
\begin{array}{ll}
    \label{auxilary}
    \tilde{\mathcal{L}} &=
    \bar{q} \left[ i \gamma^{\mu}\nabla_{\mu} + \mu \gamma^{0} \right] q
    - \bar{q}\left(\sigma+i\gamma^5\vec{\tau}\vec{\pi}\right)q
    - \frac3{2G_1}(\sigma^2+\vec{\pi}^2) - {}\\
    &- \frac3{G_{2}} \Delta^{*b} \Delta^b -
    \Delta^{*b}\left[i q^t \textit{C} \varepsilon \epsilon^{b} 
\gamma^{5} q\right]
    -\Delta^{b}\left[i \bar{q} \varepsilon \epsilon^{b} \gamma^{5} \textit{C}
\bar{q}^t\right].
\end{array}
\end{equation}
The Lagrangians~(\ref{lagrangian}) and~(\ref{auxilary}) are
equivalent, as can be seen by using the equations of motion for the boson fields
$$
    \Delta^b=-\frac{G_2}3\,i q^t \textit{C} \varepsilon \epsilon^{b}
\gamma^{5} q\, \qquad \sigma=-\frac{G_1}3\,\bar{q}q, \qquad
    \vec{\pi}=-\frac{G_1}3\,\bar{q}i\gamma^5\vec{\tau}q.
$$
The fields $\sigma$ and $\vec{\pi}$ are color singlets, and
$\Delta^b$ is a color anti-triplet and flavor singlet. Therefore, if
$<\sigma>\;\neq\;0$, the chiral symmetry is broken dynamically, and if
 $<\Delta^b>\;\neq\;0$, the color symmetry is broken.
 
\section{Effective action}
 
Following~\cite{Ebert_Zhukovsky}, we obtain the partition function of
the theory
\begin{equation}
    Z=\int[dq][d\bar{q}][d\sigma][d\vec{\pi}][d\Delta^{*b}][d\Delta^{b}]\,
        \exp\biggl\{\,i\int d^D x\sqrt{-g}\,\tilde{\mathcal{L}}\biggr\},
\end{equation}
where $g=\det|\,g_{\mu\nu}|$. In what follows, it is convenient to
consider the effective action for boson fields, which is expressed through the
integral over quark fields
\begin{equation}
    \exp\left\{\,iS_{\rm eff}(\sigma,\vec\pi, \Delta^b, \Delta^{*b})\right\}=
        N'\int[dq][d\bar{q}]\exp\biggl\{\,i\int d^Dx\sqrt{-g}\tilde{\mathcal{L}}\biggr\},
\end{equation}
where
\begin{equation}
    S_{\rm eff}(\sigma,\vec{\pi}, \Delta^b, \Delta^{*b})=-\int d^Dx\sqrt{-g}
    \left[
           \frac{3(\sigma^2+\vec{\pi}^2)}{2G_1}+\frac{3\Delta^b\Delta^{*b}}{G_2}
    \right] + S_q,
\end{equation}
$N'$ is a normalization constant. The quark contribution to the
partition function is given by the expression
\begin{equation}
\begin{array}{lr}
    Z_q = e^{\,iS_q}=N'\int[dq][d\bar{q}]\exp\biggl(&i\int\ d^Dx\sqrt{-g}
        \Bigl[\bar{q}
                \left(
                    i\gamma^{\mu}\nabla_{\mu}-\sigma-i\gamma^5\vec{\pi}\vec{\tau}+\mu\gamma^0
                \right)q - {}\\
              &-\bar{q}
                \left(
                    i\Delta^b\varepsilon\epsilon^b\gamma^5\textit{C}
                \right)\bar{q}^t -
                q^t\left(
                    i\Delta^{*b}\textit{C}\varepsilon\epsilon^b\gamma^5
                \right)q
        \Bigr]\biggr).
\label{genfunc}
\end{array}
\end{equation}
In the mean field approximation, the fields
$\sigma$, $\vec{\pi}$, $\Delta^b$,  $\Delta^{*b}$ can be replaced by
their groundstate averages: $<\sigma>$, $<\vec{\pi}>$,
$<\Delta^b>$ and $<\Delta^{*b}>$, respectively.  
Let us choose the following ground state of our model:
$<\Delta^1>\;=\;<\Delta^2>\;=\;<\vec{\pi}>\;=\;0$ and $<\sigma>$,
$<\Delta^3> \; \neq\;0$, denoted by letters
$\sigma,\;\Delta$. Evidently, this choice breaks the color
symmetry down to the residual group $SU_c(2)$.
 
Let us find the effective potential of the model with the global minimum point that
will determine the quantities $\sigma$ and $\Delta$. By definition
$S_{\rm eff}=-V_{\rm eff}\int d^Dx \sqrt{-g}$, where
\begin{equation}
    V_{\rm
    eff}=\frac{3\sigma^2}{2G_1}+\frac{3\Delta\Delta{*}}{G_2}+\tilde V_{\rm eff};\quad
   \tilde V_{\rm eff}=-\frac{ S_q}{v},\quad v=\int d^Dx\sqrt{-g}.
\end{equation}
In virtue of the assumed vacuum structure ($<\Delta^{1,\,2}>=0$), the
functional integral for $S_q$ in~(\ref{genfunc}) factorizes into two parts
\begin{equation}
\begin{array}{lr}
    \label{zq}
    Z_q&=\exp{iS_q(\sigma,\Delta)}=N'\int[d\bar{q}_3][dq_3]
    \exp{\left(i\int d^D x \sqrt{-g}\;\bar{q_3}\tilde{D}q_3\right)}\times {}\\
    &\times\int\left[d\bar{Q}\right]\left[dQ\right]
    \exp{\left(i\int d^D x\sqrt{-g}
        \left[
           \bar{Q}\tilde{D}Q+\bar{Q}M\bar{Q}^t+Q^t\bar{M}Q
        \right]\right)},
\end{array}
\end{equation}
where  $q_3$ is the quark field of color 3,
and $Q\equiv(q_1,\;q_2)^t$ is the doublet of quarks of colors 1,2. The
following notations are also introduced
\begin{equation}
\tilde{D}=i\gamma^{\mu}\nabla_{\mu}-\sigma+\mu\gamma^0,\, 
\quad \bar{M}=-i\Delta^{*}\textit{C}\varepsilon\tilde{\epsilon}\gamma^5,
\quad M=-i\Delta\varepsilon\tilde{\epsilon}\gamma^5\textit{C}.
\end{equation}
Here
$\tilde{\epsilon}\equiv(\tilde{\epsilon})^{\alpha\beta}$ denotes an
antisymmetric tensor in the color subspace corresponding to the
$SU_c(2)$ group.
 
Clearly, integration over  $q_3$ in ~(\ref{zq}) leads to $\Det\tilde{D}$.
Let us next define  $\Psi=(Q^t,\bar{Q})$ and introduce a matrix operator
$$
Z= \left(
  \begin{array}{cc}
    2\bar{M}, & -\tilde{D}^t \\
    \tilde{D}, & 2M \\
  \end{array}
\right).
$$
Then the Gaussian integral in~(\ref{zq}) can be written in a compact form
\begin{equation}
\label{gauss} \int\left[d\Psi\right] e^{\frac i2 \int d^D x
\sqrt{-g}\;\Psi^t Z \Psi}=\Det\!^{1/2} Z.
\end{equation}
Using the general formula
$$
    \Det\left(
          \begin{array}{cc}
            A & B \\
            \bar{B} & \bar{A} \\
          \end{array}
        \right)
    =\Det\left[-\bar{B} B + \bar{B}A\bar{B}^{-1}\bar{A}\right]
    =\Det\left[\bar{A}A-\bar{A}B\bar{A}^{-1}\bar{B}\right],
$$
we obtain
\begin{equation}
\begin{array}{llll}
\label{effaction}
    \exp\left(i S_q(\sigma,\Delta) \right)
        &=& & N'\Det\left(\tilde{D}\right)\,\Det\!^{1/2}\left[4M\bar{M}+M\tilde{D}^t
            M^{-1}\tilde{D}\right]={}\\
        &=&& N'\Det\left[(i\hat{\nabla}-\sigma+\mu\gamma^0)\right]\times{}\\
    & &&\times\Det\!^{1/2}\left[4|\Delta|^2 + (-i\hat{\nabla}-\sigma+\mu\gamma^0 )
        (i\hat{\nabla}-\sigma+\mu\gamma^0 ) \right].
\end{array}
\end{equation}
Here, the first determinant is over spinor, flavor and coordinate
spaces, and the second one is over the two-dimensional color space as
well, and $\hat{\nabla}=\gamma^\mu\nabla_\mu$.
 
\section{Evaluation of determinants in the static Einstein universe}
 
Let us first calculate the contribution of the determinant
$$
\begin{array}{lll}
    \ln\Det\left[(i\hat{\nabla}-\sigma+\mu\gamma^0)\right]&=&
    {1\over 2}\ln\Det\left[(i\hat{\nabla}-\sigma+\mu\gamma^0)\right]+
   {1\over 2} \ln\Det\left[\gamma^5(i\hat{\nabla}-\sigma+\mu\gamma^0)\gamma^5\right]=\\
    & =&{1\over 2}\ln\Det\left[\gamma^\mu\gamma^\nu\nabla_{\mu}\nabla_{\nu}-2i\mu\nabla_0
    -\mu^2+\sigma^2\right].
\end{array}
$$
In our case, $\Gamma_0$ is evidently equal to zero, and
$g^{\mu\nu}\nabla_\mu\nabla_\nu=\frac{\partial^2}{\partial
t^2}-\nabla^2$, where
$\nabla^2=g^{ij}\nabla_i\nabla_j$
is the
spinor Laplacian. Then
\begin{equation}
    2\ln\Det\left[(i\hat{\nabla}-\sigma+\mu\gamma^0)\right]=
    \ln\Det\Bigl[\left(\frac\partial{\partial t}-i\mu\right)^2
    -\nabla^2+\frac14R+\sigma^2\Bigr].
\label{above}
\end{equation}
 
In (\ref{above}) we used the identity \cite{Parker_Toms} 
\begin{equation}
\label{identity}
    \gamma^\mu\gamma^\nu\nabla_{\mu}\nabla_{\nu}=g^{\mu\nu}\nabla_\mu\nabla_\nu+\frac14R,
\end{equation}
where $R$ is the scalar curvature of the space-time. 
Next, consider the contribution of the second determinant
\begin{equation}
\begin{array}{ll}
    \label{det2}
    & \Det\left[4|\Delta|^2 + (-i\hat{\nabla}-\sigma+\mu\gamma^0 )
        (i\hat{\nabla}-\sigma+\mu\gamma^0 ) \right]=\\
    &    =\Det\left[4|\Delta|^2+\gamma^\mu\gamma^\nu\nabla_{\mu}\nabla_{\nu}
        +\sigma^2+\mu^2-2\mu\sigma\gamma^0+2i\mu\gamma^0\gamma^k\nabla_k\right].
\end{array}
\end{equation}
In analogy with the case of a flat Minkowski space-time, consider the Hamiltonian of a
massless particle
\begin{equation}
\label{H0}
    \hat{H}=\vec{\alpha}\,\hat{\vec{p}},
\end{equation}
where $\alpha^k=\gamma^0\gamma^k$, and
$\left(\hat{p}\right)^k=-i\nabla_k,\;k=1,\ldots, D-1$. 
One can easily
demonstrate that in the case of the static  metric~(\ref{metric})
\begin{equation}
\label{H02}
 \hat{H}^2=-\nabla^2+\frac14R.
\end{equation}
With consideration of~(\ref{identity}) and~(\ref{H02}),
formula~(\ref{det2}) can be rewritten in the form
\begin{equation}
\begin{array}{ll}
    & \Det\left[4|\Delta|^2 + (-i\hat{\nabla}-\sigma+\mu\gamma^0)
        (i\hat{\nabla}-\sigma+\mu\gamma^0 ) \right]=\\
    & =\Det\left[4|\Delta|^2
        +\sigma^2+\mu^2-\hat p_0^2+\hat{H}^2-2\mu\left(\hat{H}+
        \sigma\gamma^0\right)\right],
\end{array}
\end{equation}
where $\hat p_0=i\partial_0$.
Finally, let us introduce the Hamiltonian of a massive particle
\begin{equation}
    \label{H}
    \hat{\mathcal H} =\vec{\alpha}\,\hat{\vec{p}}+\sigma\gamma^0,
\end{equation}
where $\hat{\vec{p}}=-i\vec \nabla$. It is related to the massless
operator as follows:
\begin{equation}
    \hat{\mathcal H}^2 =\hat{H}^2+\sigma^2.
\end{equation}
Then,~(\ref{det2}) is rewritten in the form
\begin{equation}
\begin{array}{ll}
    & \Det\left[4|\Delta|^2 + (-i\hat{\nabla}-\sigma+\mu\gamma^0)
        (i\hat{\nabla}-\sigma+\mu\gamma^0 ) \right]=\\
    & =\Det\left[4|\Delta|^2+
        \mu^2-\hat{p}_0^2+\hat{\mathcal H}^2-2\mu\hat{\mathcal H}\right].
\end{array}
\end{equation}
 
Thus, in order to calculate the determinants defining the effective action, one has to find the
eigenvalues of the operators $\hat{H}$ and
$\hat{\mathcal{H}}$. In particular, they may be found exactly for the
case of the static D-dimensional Einstein universe. They are expressed
through the corresponding eigenvalues of the Dirac operator on the sphere
$\mathbb{S}^{D-1}$~\cite{Camporesi, Weinberg}. The line element
\begin{equation}
   ds^2=dt^2-a^2(d\theta^2+\sin^2\theta d\Omega_{D-2})
\end{equation}
gives the global topology $\mathbb{R}\otimes\mathbb{S}^{D-1}$ of the
universe, where $a$ is the radius of the universe, related to the
scalar curvature by the relation $R=(D-1)(D-2)a^{-2}$. The volume of
the universe is determined by the formula
\begin{equation}
\label{volume}
 V(a)=\frac{2\pi^{D/2}a^{D-1}}{\Gamma(\frac D2)}.
\end{equation}
 
Let us denote the absolute values of the eigenvalues of the operators~(\ref{H0}) and
~(\ref{H}) by $\omega_l$ and $E_l$ respectively. Then, according to \cite{Camporesi, Weinberg}
one has
\begin{equation}
    \begin{array}{llllllll}
        \hat{H}\,\psi_l&=&\pm\,\omega_l\psi_l,  & \omega_l&=&\frac{1}{a}\left(l+\frac{D-1}{2}\right)& &\\
        \hat{\mathcal H}\,\psi_l&=&\pm\, E_l\psi_l, &
        E_l&=&\sqrt{\omega_l^2+\sigma^2},& &l=0,1,2\ldots \; .
    \end{array}
\label{discrete}
\end{equation}
The degeneracies of  $\omega_l$ and $E_l$ are equal to
\begin{equation}
    d_l=\frac{2^{[(D+1)/2]}\Gamma(D+l-1)}{l!\Gamma(D-1)},
\end{equation}
where $[x]$ is the integer part of  $x$.
 
\section{Thermodynamic potential and gap equations}
 
Next, let us take the complete system of eigenfunctions
$\Psi_k(t,\vec{x})=e^{-i p_0 t}\psi_l(\vec{x})$ as a basis.
Then the
operators in the two determinants are diagonal and the latter are reduced to
products of corresponding eigenvalues
$E_l^2-(p_0-\mu)^2$ and $4|\Delta|^2-p_0^2+(E_l\pm \mu)^2$. Evidently,
there appears a gap in the spectrum of the second operator
proportional to $|\Delta|$, and as a result, the diquark condensate is
formed. Using the relation $\det O=\exp(\tr\ln O)$ and following the
standard procedure, one obtains the following expression for the
quark contribution to the effective potential
\begin{equation}
\begin{array}{ll}
    \label{V}
 \tilde V_{\rm eff}=-\frac{ S_q}{v}&=i \frac{N_f}{V}\int\frac{dp_0}{2\pi}\sum\limits_{l=0}^{\infty}\sum\limits_{\pm}
        d_l\Bigl\{\ln\left[(E_l\pm\mu)^2-p_0^2\right]+\\
    &      +2\ln\left[4|\Delta|^2-p_0^2+(E_l\pm\mu)^2\right]\Bigl\},
\end{array}
\end{equation}
where $v=Vt$ and $V$ is the volume of the universe, defined in~(\ref{volume}),
$t$ is the time interval, and $N_f$ denotes the number of flavors $N_f=2$.
 
In the case of finite temperature $T=1/\beta>0$, 
the following substitutions in~(\ref{V}) should be made:
$$
    \int\frac{dp_0}{2\pi}(\cdots)\rightarrow i T\sum_n(\cdots),\quad
    p_0\rightarrow i\omega_n,\quad
    \omega_n=\frac{2\pi}{\beta}\left(n+\frac12\right),
    \quad n=0,\pm1\pm2,\ldots,
$$
where $\omega_n$ is the Matsubara frequency. Then the quark
contribution to the effective potential (\ref{V}) becomes the
thermodynamic potential $\Omega_q$ 
\begin{equation}
\begin{array}{ll}
    \label{VT}
     \Omega_q & = -\frac{N_f}{V\beta}\sum\limits_{n=-\infty}^{n=\infty}\sum\limits_{l=0}^{\infty}\sum\limits_{\pm}
        d_l\Bigl\{\ln\left[(E_l\pm\mu)^2+\omega_n^2\right]+\\
    &      +2\ln\left[4|\Delta|^2+\omega_n^2+(E_l\pm\mu)^2\right]\Bigl\}.
\end{array}
\end{equation}
The summation over the Matsubara frequencies can now be performed with
the use of the formula
\begin{equation}
\label{matsubara}
 \sum_n \ln\left(\omega_n^2+\rho^2\right)=\sum_n
\int\limits_{1/\beta^2}^{\rho^2}d x^2
\frac{1}{\omega_n^2+x^2}+\sum_n\ln\left(\omega_n^2+\frac{1}{\beta^2}\right),
\end{equation}
where, according to~(\ref{VT}),  $\rho$ denotes
$\sqrt{(E_l\pm\mu)^2}$ or $\sqrt{4|\Delta|^2+(E_l\pm\mu)^2}$,
respectively.
In the following, we shall neglect the last term 
in~(\ref{matsubara}),
since it is independent of $\sigma$ and
$\Delta$,
and
the first term 
can be rewritten as
\begin{equation}
    \sum_n \int\limits_{1/\beta^2}^{\rho^2}d x^2 \frac{1}{\omega_n^2+x^2}=
    2\ln\cosh(\rho\beta/2)+{\rm
    const}=\rho\beta+2\ln\left(1+e^{-\rho\beta}\right)+{\rm const}.
\end{equation}
Then the final form of the thermodynamic potential looks like
\begin{equation}
\begin{array}{ll}
    \label{potential}
    & \Omega(\sigma,\Delta)=3\left(\frac{\sigma^2}{2G_1}+\frac{|\Delta|^2}{G_2}\right)-\\
    &  -\frac{N_f}{V}(N_c-2)\sum\limits_{l=0}^{\infty} d_l\left\{ E_l+T\ln\left(1+e^{-\beta(E_l-\mu)}\right)+
       T\ln\left(1+e^{-\beta(E_l+\mu)}\right)\right\}-\\
    &  -\frac{N_f}{V}\sum\limits_{l=0}^{\infty} d_l\biggl\{\sqrt{(E_l-\mu)^2+4|\Delta|^2}
            +\sqrt{(E_l+\mu)^2+4|\Delta|^2}+\\
    &          +2T\ln\left(1+e^{-\beta\sqrt{(E_l-\mu)^2+4|\Delta|^2}}\right)+
                2T\ln\left(1+e^{-\beta\sqrt{(E_l+\mu)^2+4|\Delta|^2}}\right)\biggl\}.
\end{array}
\end{equation}
 
In order to find the correspondence of the result obtained with the
case of the flat space-time \footnote{Due to different topologies of the flat space-time and the
Einstein universe, the expressions for their line
elements do not relate to each other by any simple limiting
procedure.}, we consider the zero curvature limit,
i.e., $R\to 0\,\,(a\to \infty), $ of the expression for
the thermodynamic potential. In this limit
\[
R\to 0,\,a\to\infty, l\to\infty,\,\omega_l\to la^{-1},\,\sum_l\to \int
dl=a\int d\omega_l=a\int dp
\]
(where we have changed the variables $\omega_l\to p$) and, since
\[
{\Gamma(D+l-1)\over \Gamma (l+1)}\approx l^{D-2},
\]
 the degeneracy is equal to
\[
d_l\,=\,{2^{[(D+1)/2]}\Gamma (D+l-1)\over \Gamma (l+1) \Gamma
  (D-1)}\approx {2^{[(D+1)/2]}\over (D-2)!}l^{D-2}.
\]
Then, taking expression (\ref{volume}) for the volume $V(a)$ into consideration, we finally
obtain
\begin{equation}
\begin{array}{ll}
& \Omega=3\left({\sigma^2\over 2G_1}+{\Delta^2\over G_2}\right)-\\
& -2N_f\int{ d^3p\over (2\pi)^3}\left\{
  E_p+T\ln\left(1+e^{-\beta(E_p-\mu)}\right)+
       T\ln\left(1+e^{-\beta(E_p+\mu)}\right)\right\}-\\
 &  -2N_f\int {d^3p\over (2\pi)^3}\biggl\{\sqrt{(E_p-\mu)^2+4|\Delta|^2}
            +\sqrt{(E_p+\mu)^2+4|\Delta|^2}+\\
    &          +2T\ln\left(1+e^{-\beta\sqrt{(E_p-\mu)^2+4|\Delta|^2}}\right)+
                2T\ln\left(1+e^{-\beta\sqrt{(E_p+\mu)^2+4|\Delta|^2}}\right)\biggl\}.
\end{array}
\end{equation}
This equation 
coincides with the corresponding result for the
flat space-time (see, e.g., \cite{Alford,Berges,Schwarz}).
 
Now, imposing the condition on the effective potential, $\Omega(0,0)=0$, we
  should subtract a corresponding constant from it. The thermodynamic potential,
  normalized in this way,  is still divergent at large $l$,
  and hence, we should introduce a (soft) cutoff in the summation over $l$
  by means of the multiplier $e^{-\omega_l/\Lambda}$, where $\Lambda$
  is the cutoff parameter.
 
As is known, in the flat space-time, the 
regularization cutoff constant $\Lambda$ can be determined from the
experimental results in measurements of the pion 
mass or the pion
decay constant. However, in the case of a curved space-time, there are no
experiments of this kind, and in principle we cannot  determine the
value of the cutoff parameter. Therefore, similar to \cite{Huang_Hao_Zhuang}, we shall
concentrate primarily on a qualitative discussion of the results
obtained, concerning the phase structure of the universe.
 
To this end, we shall divide all the dimensional quantities that enter
the thermodynamic potential by the corresponding power of $\Lambda$ to
make them dimensionless, i.e., 
$\Omega/\Lambda^D, \;
\sigma/\Lambda, \;
\Delta/\Lambda, \; \Lambda^{D-2}G_{1,\,2}, \; \Lambda^{D-1}V, \;
R/\Lambda^2, \; T/\Lambda, \; \mu/\Lambda, \; \omega_l/\Lambda$, and
denote them using
the same letters as before: 
$\Omega, \;
 \sigma, \; \Delta, \; G_{1,\,2}, \; V, \; R,
\; T, \; \mu, \; \omega_l$. Then the regularized thermodynamic
potential is written as
\begin{equation}
\begin{array}{ll}
   \label{potential1}
    & \Omega^{\rm reg}(\sigma,\Delta)=3\left(\frac{\sigma^2}{2G_1}+\frac{|\Delta|^2}{G_2}\right)-\\
    &  -\frac{N_f}{V}(N_c-2)\sum\limits_{l=0}^{\infty} e^{-\omega_l} d_l\left\{ E_l+T\ln\left(1+e^{-\beta(E_l-\mu)}\right)+
       T\ln\left(1+e^{-\beta(E_l+\mu)}\right)\right\}-\\
    &  -\frac{N_f}{V}\sum\limits_{l=0}^{\infty} e^{-\omega_l} d_l\biggl\{\sqrt{(E_l-\mu)^2+4|\Delta|^2}
            +\sqrt{(E_l+\mu)^2+4|\Delta|^2}+\\
    &          +2T\ln\left(1+e^{-\beta\sqrt{(E_l-\mu)^2+4|\Delta|^2}}\right)+
                2T\ln\left(1+e^{-\beta\sqrt{(E_l+\mu)^2+4|\Delta|^2}}\right)\biggl\}.
\end{array}
\end{equation}
 
In the following Section, we shall perform
the numerical calculation of the
points of the global minimum of the finite regularized thermodynamic
potential $\Omega^{\rm reg}(\sigma,\Delta)-\Omega^{\rm reg}(0,0)$ (they should of course
coincide with the minima of the potential
$\Omega^{\rm reg}(\sigma,\Delta)$), and with the use of them, consider
phase transitions in the Einstein universe. To this end, we have to
consider the following gap equations:
\begin{equation}
{\partial \Omega^{\rm reg}\over \partial \sigma }=0,\,\,{\partial
  \Omega^{\rm reg}\over \partial \Delta }=0.
\label{gap}
\end{equation}
 
\section{Phase transitions }
 
As it is seen from~(\ref{potential}), the thermodynamic potential
depends on two independent coupling constants $G_1$ and
$G_2$. In what follows, we shall fix the constant $G_2$, similarly to
what has been done in the flat
case~\cite{Berges,Ebert_Zhukovsky}, by using the relation
\begin{equation}
    G_2=\frac38G_1.
\end{equation}
For numerical estimates, let us further choose
the constant $G_1$ such that the
chiral and/or color symmetries were completely broken. For
comparison, we shall consider two cases of the choice of the constant:
1) $G_1=10$ (this value formally 
corresponds to that of \cite{Huang_Hao_Zhuang}) and 2) $G_1=20$
~\footnote
{Different choices of the value of $G_1$ do not change 
the results in principle, but the scales of parameters are changed
significantly (see below).}. 
Moreover, let us now limit ourselves to the investigation 
of the case $D=4$ only.
 
Let us first consider phase transitions 
in the case of zero
temperature. Then, the thermodynamic potential takes the form
\begin{equation}
\begin{array}{ll}
     \label{zeropotantial}
    & \Omega^{\rm reg}(\sigma,\Delta)=3\left(\frac{\sigma^2}{2G_1}+\frac{|\Delta|^2}{G_2}\right)-\\
    &  -\frac{N_f}{V}(N_c-2)\sum\limits_{l=0}^{\infty} e^{-\omega_l} d_l\left\{E_l+(\mu-E_l)\theta(\mu-E_l)\right\} -\\
    &  -\frac{N_f}{V}\sum\limits_{l=0}^{\infty} e^{-\omega_l} d_l\left\{\sqrt{(E_l-\mu)^2+4|\Delta|^2}+\sqrt{(E_l+\mu)^2+4|\Delta|^2}\right\}.
\end{array}
\end{equation}
 
\begin{figure}[h]
   \noindent
  \centering
 $
\begin{array}{cc}
 \includegraphics[width=0.45\textwidth]{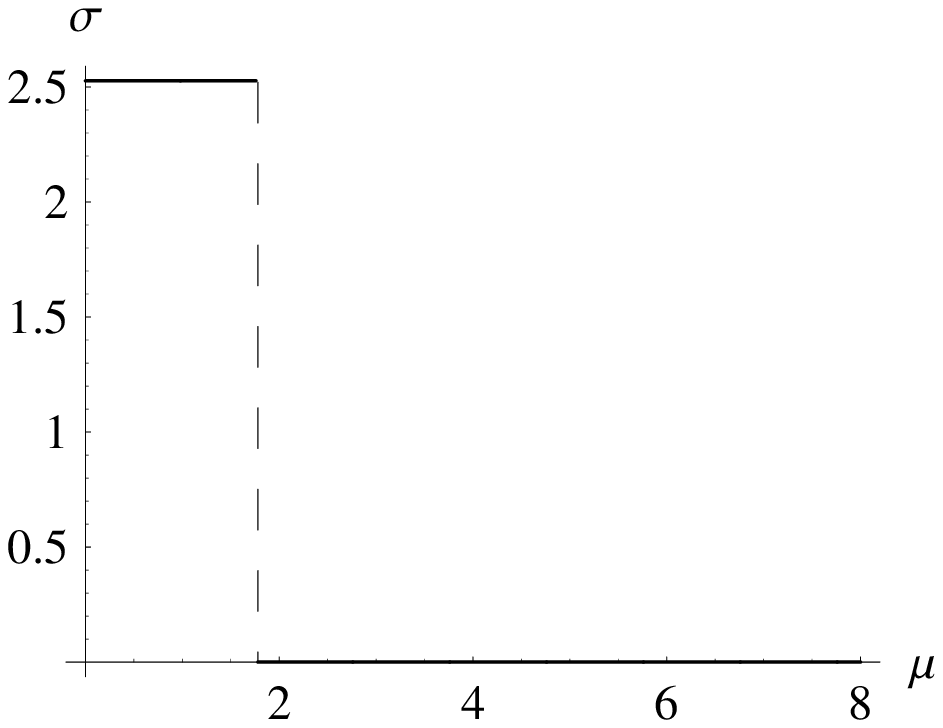} &
\includegraphics[width=0.45\textwidth]{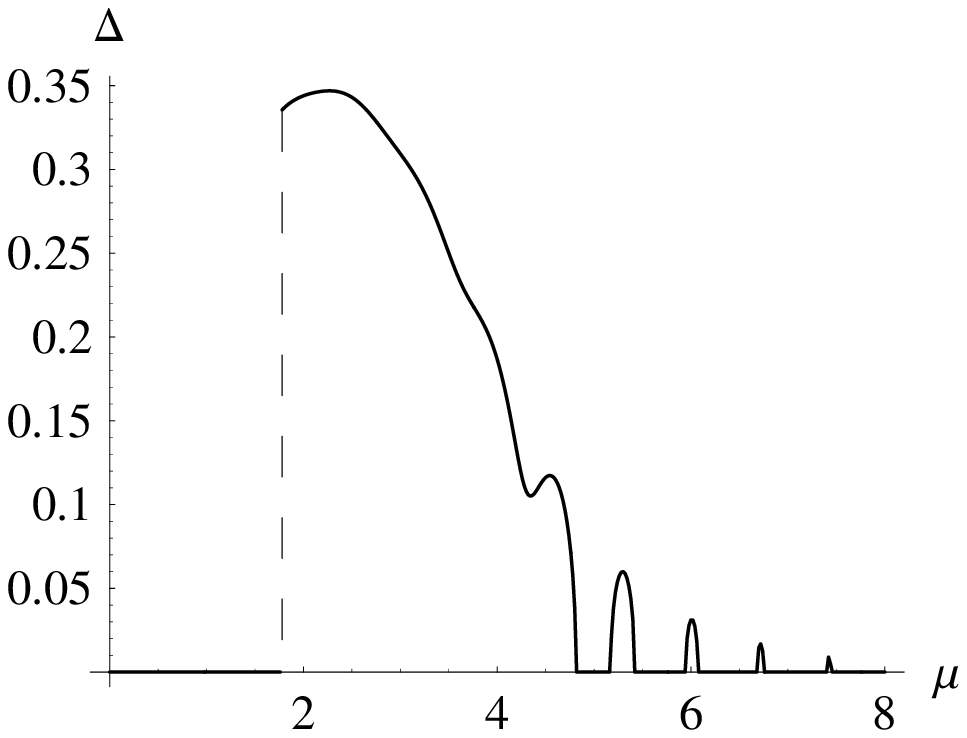}
  \end{array}
    $
  \caption{Condensates $\sigma$
and
$\Delta$ as functions of
$\mu$ for $R=3$, $G_1=10$,
$T=0$
 (all quantities are given
in units of $\Lambda$).}
   \label{Cond1003}
\end{figure}
\vspace{0.5cm}
\begin{figure}[h]
   \noindent
  \centering
 $
\begin{array}{cc}
 \includegraphics[width=0.45\textwidth]{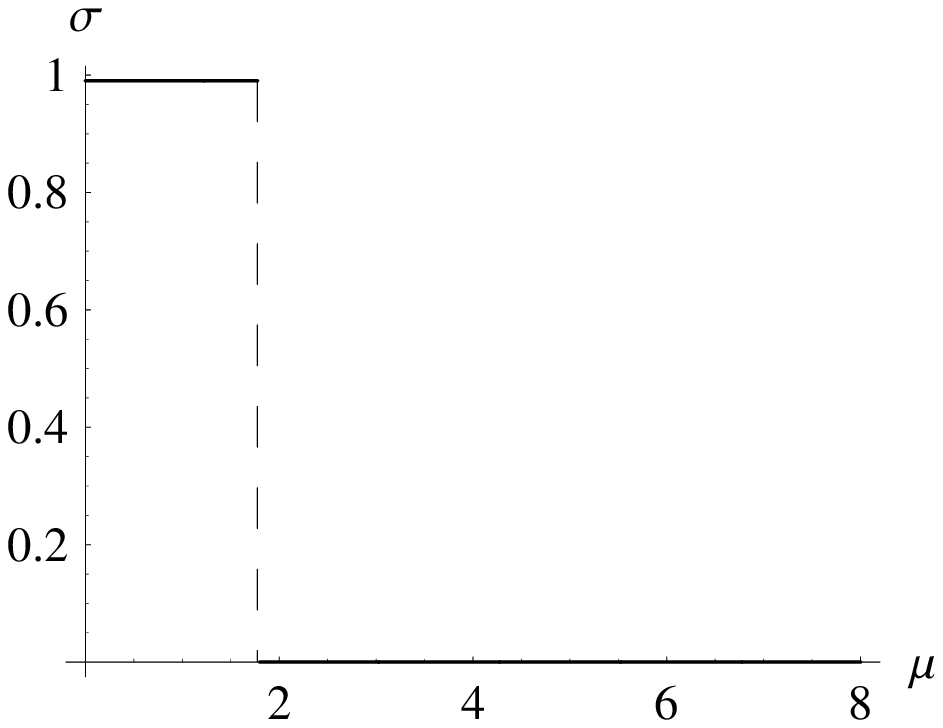} &
\includegraphics[width=0.45\textwidth]{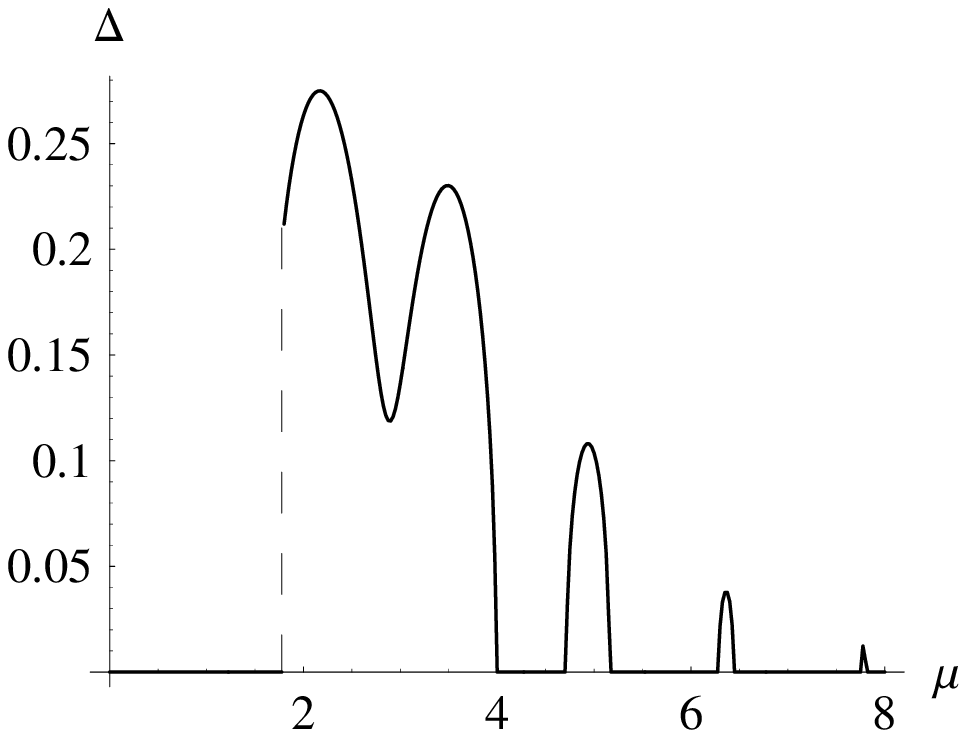}
  \end{array}
    $
  \caption{Condensates $\sigma$
and
$\Delta$ as functions of
$\mu$ for $R=12$, $G_1=10$,
$T=0$
 (all quantities are given
in units of $\Lambda$).}
   \label{Cond10012}
\end{figure}
\vspace{0.5cm}
\begin{figure}[h]
   \noindent
  \centering
 $
\begin{array}{cc}
 \includegraphics[width=0.45\textwidth]{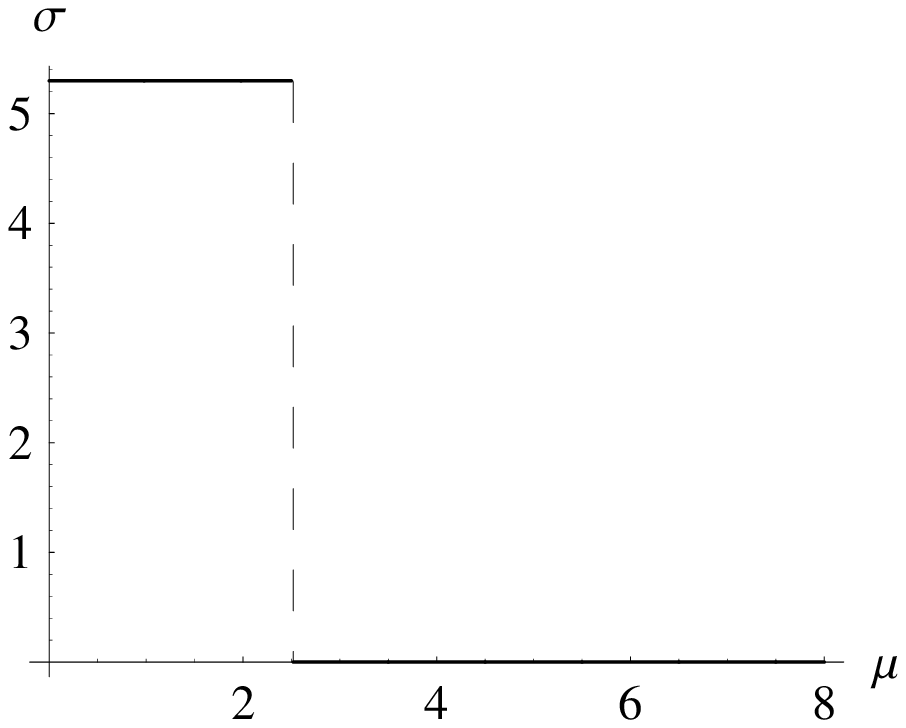} &
\includegraphics[width=0.45\textwidth]{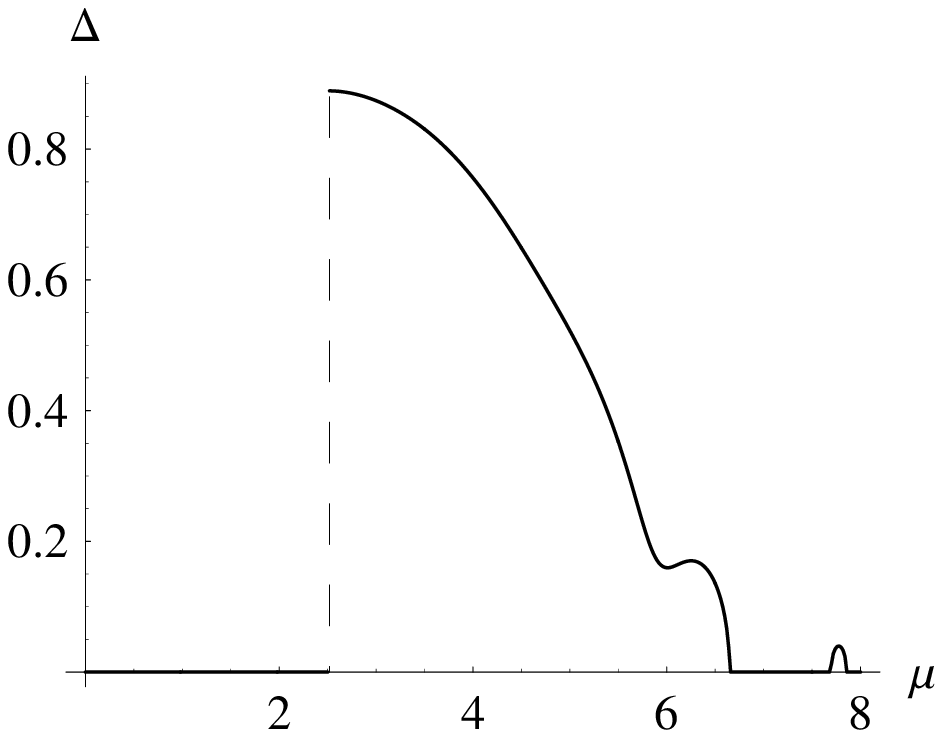}
  \end{array}
    $
  \caption{Condensates $\sigma$
and
$\Delta$ as functions of
$\mu$ for $R=12$, $G_1=20$,
$T=0$
 (all quantities are given
in units of $\Lambda$).}
   \label{Cond20012}
\end{figure}
Figs.~\ref{Cond1003}, \ref{Cond10012}, 
\ref{Cond20012}
show the  behavior  of the point of the global minimum 
$(\sigma,\Delta)$, determined from
the gap equations (\ref{gap}) for
the potential~(\ref{zeropotantial}), as a
function of $\mu$ at different values of $R=3, R=12$ and $T=0$, for
$G_1=10$ and $G_1=20$.

\begin{figure}[h]
\noindent
\centering{
\includegraphics[width=0.55\textwidth]{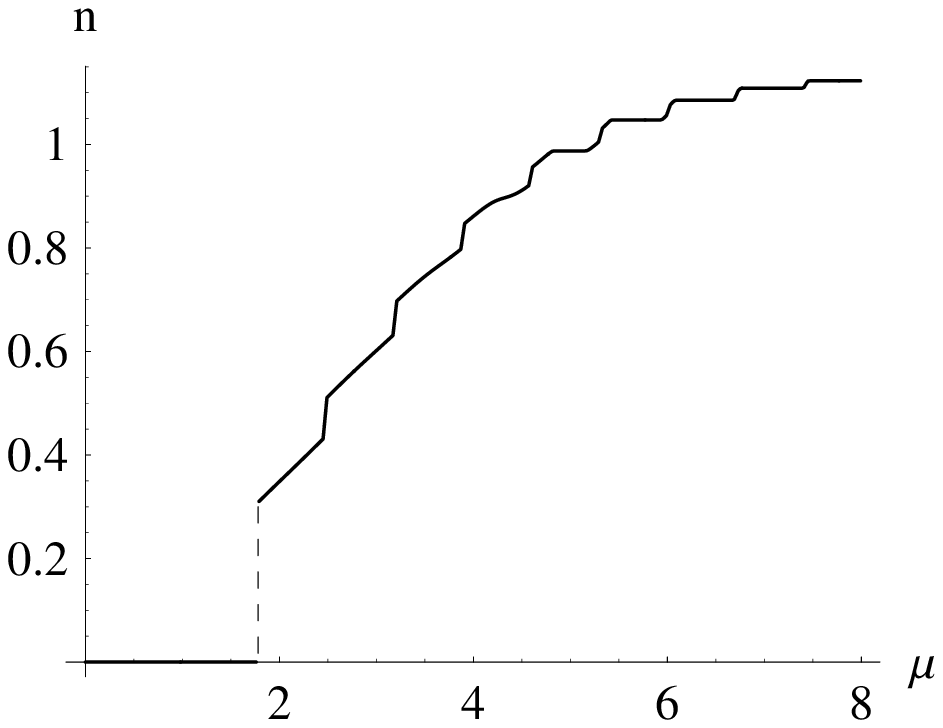}
} \caption{The fermion number density at R=3, T=0, $G_1=10$.}
 \label{density0} 
\end{figure}

Using the formulas for the thermodynamic potential
$\Omega^{\rm reg}$,
we can calculate the fermion number density in the system
\begin{equation}
    n=-\frac{\partial \Omega^{\rm reg}}{\partial \mu} \quad\mbox{at fixed } T.
\end{equation}
The function  $n(\mu)$ at fixed $R$ is depicted in Fig.~\ref{density0}.
The step-wise character of the curve is explained by the discreteness
of the energy levels and by the presence of the step-function in
~(\ref{zeropotantial}).
It is seen from these figures 
that, when the chemical potential 
exceeds the critical value  $\mu_c$, the chiral symmetry is
restored and the color symmetry is broken\footnote{It should be
 mentioned that for $R=3$, $G_1=10$ our value for $\mu_c\approx 1.75$
 is close to that from \cite{Huang_Hao_Zhuang}.}. 

One may ask, why
  the diquark condensate $\Delta$ vanishes for large chemical
  potential $\mu$,
i.e., why CSC will be destroyed at large densities? 
(An analogous behavior is seen also in Fig.6 of  \cite{Schwarz}.) 
In fact, since in the weak coupling QCD the nonvanishing CSC phase is
present for arbitrary 
large values of $\mu$, the NJL-results for large values of
the chemical potential could be an artefact of the model and should be
considered with care.
 
As is clear from Figs.~\ref{Cond1003}, \ref{Cond10012}, 
\ref{Cond20012}, \ref{density0}, 
a first order 
phase transition takes place at the point $\mu=\mu_c$ (the first
derivative of the 
themodynamical 
potential is discontinuous).
 
\begin{figure}[h]
\noindent
\centering
$
\begin{array}{cc} 
\includegraphics[width=0.45\textwidth]{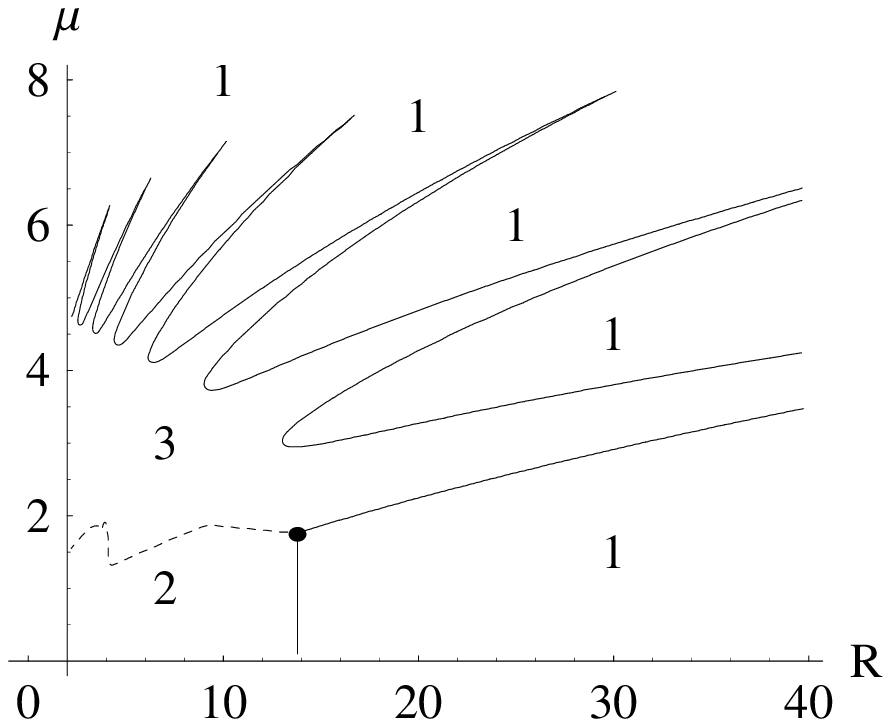}&
 \psfrag{Mu}{$\mu$}
\includegraphics[width=0.45\textwidth]{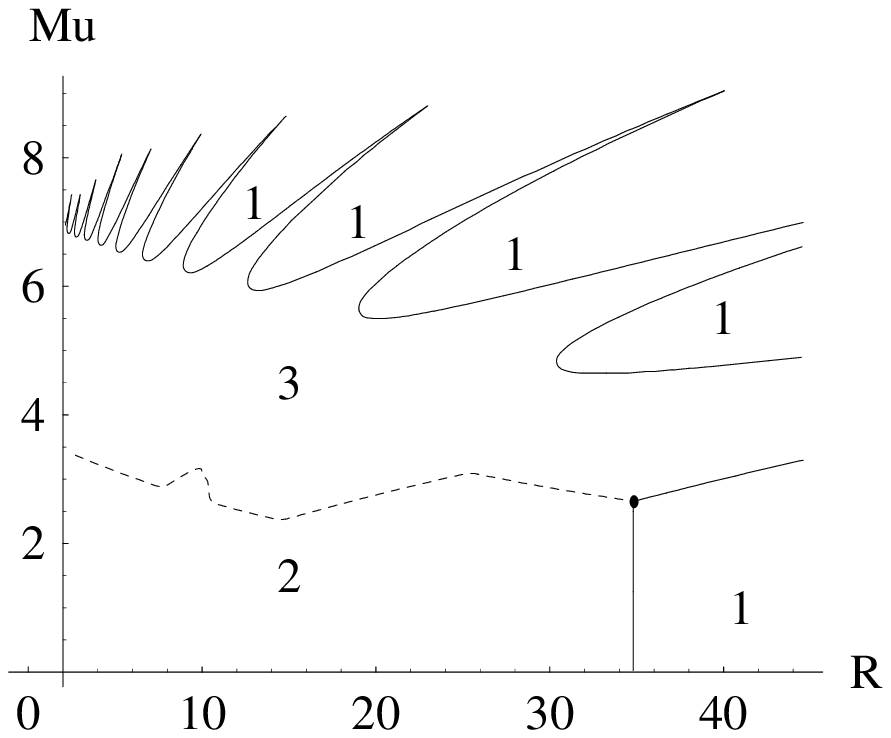}
\end{array}
$
 \caption{The phase portrait at T=0 for $G_1=10$ (left picture) and
   $G_1=20$ (right picture).
Dotted (solid) lines denote first (second) order phase transitions. The bold 
point denotes a tricritical point.
The numbers 1,2 and 3 designate the symmetric, chiral and
superconducting phases, respectively.}
\label{PP_T=0}
\end{figure}

In Fig.~\ref{PP_T=0}, the $\mu - R$-phase portrait of the system at zero
temperature 
is depicted for two choices of $G_1=10$ and $G_1=20$ (1, 2 and 3
denote the symmetric, chiral and  
superconducting phases, respectively).

For points in the symmetric
phase 1, the
global minimum of the thermodynamic potential is at $\sigma=0, \;
\Delta=0$ (chiral and color symmetries are unbroken). In the region of
phase 2, only chiral symmetry is broken and
$\sigma\neq0, \; \Delta=0$. The points in phase 3 correspond to
the formation of the diquark condensate (color superconductivity) and the
minimum takes place at
$\sigma=0, \; \Delta\neq0$. The dotted line in 
Fig. \ref{PP_T=0}
denotes the
first order phase transition, and by the solid line the second order
phase transition is depicted. The bold point denotes the tricritical
point between 
transitions of the first and second orders.
As was pointed out in~\cite{Huang_Hao_Zhuang}, the vertical line at
$T=0$ between phases  1 and 2 implies that the lowest eigenvalue of
$\omega_0$ is non-vanishing. In the vicinity of this line, a second
order phase transition takes place, i.e., in this case, the curvature
$R$ behaves exactly in the same way as the temperature  $T$ does in
the flat space. Let us mention, that, as in the flat space
(see,~\cite{Berges}), 
at $T=0$, no mixed phase with both condensates different from zero
takes place. Comparison of the plots for values $G_1=10$ and $G_1=20$
demonstrates no significant modification of the form of the phase
portrait made by the different choices
of the coupling constant. However, with growing value of $G_1$, the critical
values of $R$ and $\mu$ increase.
 
\begin{figure}[h]
\noindent
    \centering
    $
    \begin{array}{cc}  
\includegraphics[width=0.45\textwidth]{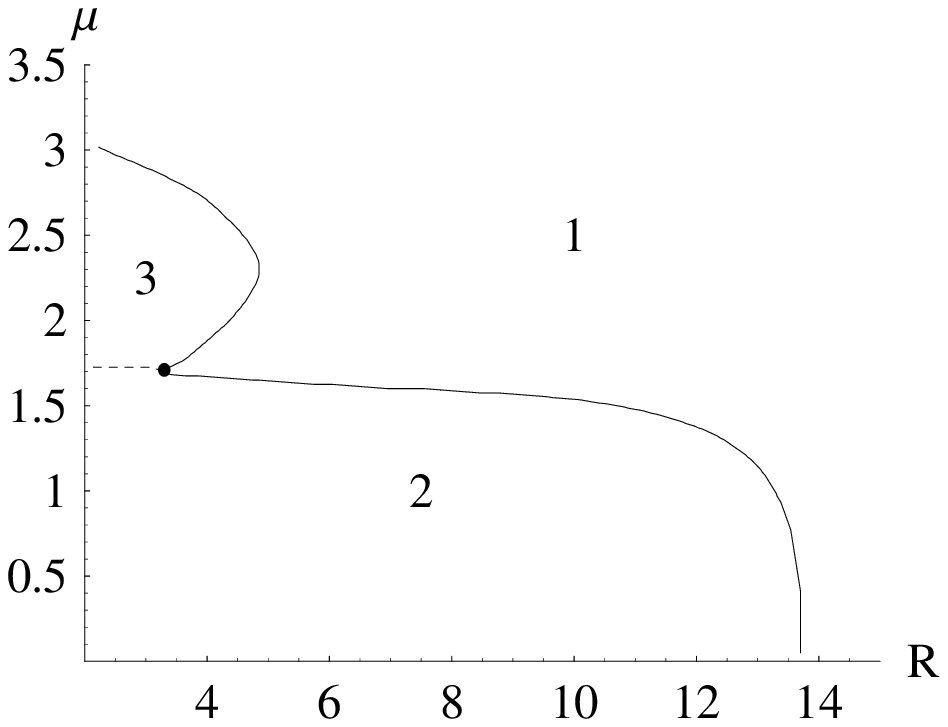}&
\includegraphics[width=0.45\textwidth]{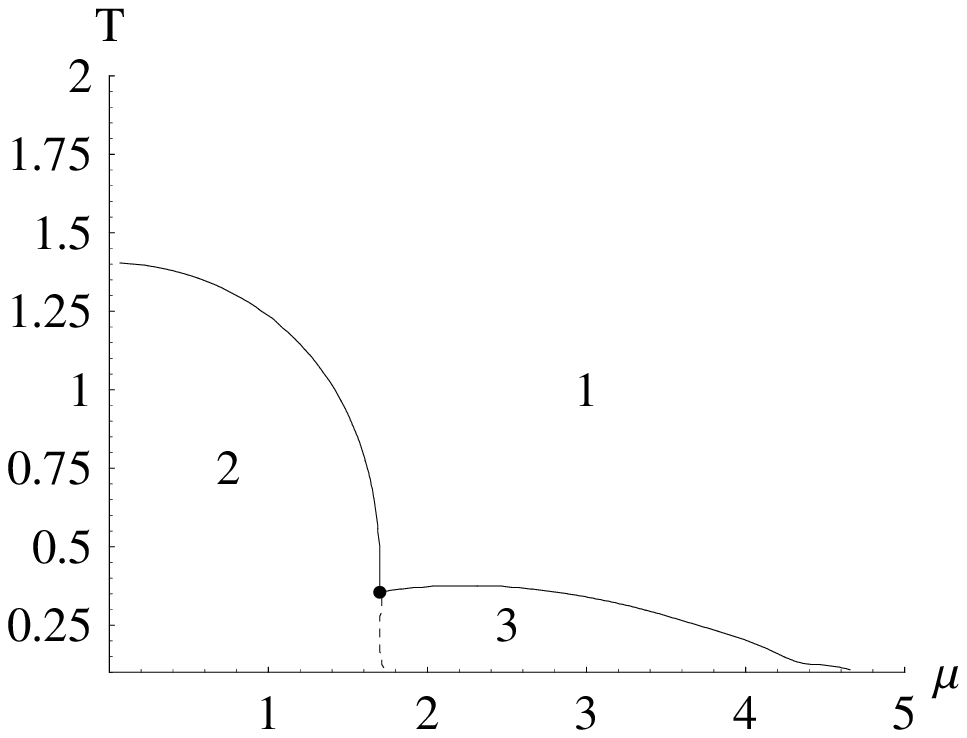} 
\end{array}
    $
 \caption{The phase portraits at T=0.35 (left picture) and at R=3 (right
   picture), $G_1=10$.} \label{PP_R5}\end{figure}

Moreover, the oscillation effect  clearly visible in the phase curves
of Fig. \ref{PP_T=0} 
should be mentioned. This  may be explained by the discreteness of the
fermion energy levels (\ref{discrete}) in the static gravitational
field. 
This effect may be compared to the similar effect in the
magnetic field $H$, where fermion levels 
are also discrete (the Landau levels).
The corresponding magnetic
oscillations of the $\mu-\sqrt eH$-phase portrait in dense cold quark
matter with four-fermion interactions
were found in paper \cite{ekvv}, where it was demonstrated that in the
massless case, such a phase structure leads unavoidably to the
standard van Alphen-de Haas magnetic oscillations of some
thermodynamical quantities, including magnetization, pressure and
particle density\footnote{
On the other hand, since these oscillations are related to the
 restoration of 
 color symmetry for large $\mu$, they seem to be here rather an
 artefact of the NJL model.}.
In addition, we considered also phase transitions at finite temperatures.
In  Fig.~\ref{PP_R5}, $\mu - R$- and $T-\mu$-
phase portraits  are depicted.
No mixed phase with simultaneous nonzero values of $\sigma\neq0, \;
\Delta\neq0$ can be seen in these plots.

It is clear from Fig.\ref{PP_R5} that with growing temperature both
the chiral and color symmetries are restored. 
The similarity
of plots in $R-\mu$ and 
$\mu-T$ axes leads one to the conclusion that
the parameters of curvature $R$ and temperature $T$ play similar roles in
restoring the symmetries of the system. 
For comparison,  we also 
studied 
the phase portrait for  the case of
vanishing curvature, 
$R=0$. 
The obtained result turns out to be very similar to the right picture in 
Fig.~\ref{PP_R5}. 
In this case,
with 
chemical potential growing from zero,  
the critical temperature for the
transition between phases 
1 and 2 decreases 
from a value slightly higher than 1.5.

\section{Conclusion}
By using a static model of curved space-time (the Einstein universe), 
we have derived a nonperturbative expression for
the effective potential 
of the Nambu--Jona-Lasinio model in the mean field approximation.
The corresponding effects of gravitation were exactly taken into account. 
This allowed
us to consider phase transitions in the vicinity of the critical
points. 
In particular, we have studied the influence of the chemical potential, the
temperature and the curvature on the chiral and color superconducting 
phase transitions. 
Moreover, 
an oscillation effect of the phase curves was  found,
which may be explained by the discreteness of the fermion energy
levels in the static gravitational field.
 
The analysis made in this paper demonstrates that the space curvature
$R$ in certain cases plays
an analogous
role as the temperature  $T$ does in
the flat space, leading to a second order phase transitions and the
restoration of 
chiral
symmetry. 
On the other hand, at certain values of the
chemical potential (and a sufficiently large coupling constant) a
superconducting phase is formed, and the color symmetry is
broken. 
However, 
as might be expected,
 the chiral and color symmetries become restored at
high temperatures and high curvatures.
Clearly, the simple schematic
model of gravitation
considered in this work may represent, at best, only
qualitative features of the diquark 
formation, 
which possibly might occur
in the cores of
heavy compact objects
like neutron stars.
Much further work has to be done towards a more quantitative
and realistic  
description of the dynamical breaking of chiral and color symmetries in 
curved space-time. This concerns, in particular, the important
question, whether  
the predicted CSC in cold heavy compact objects like neutron stars might 
eventually be destroyed by their associated strong gravitational fields.

Notice that 
our investigations
were primarily concentrated on the study of CSC in the
4-dimensional Einstein universe. It should,
however, be mentioned that the phase curves plotted
for space dimensions different from 3 
demonstrate a considerable similarity with the case studied in this work.

\section*{Acknowledgments}
One of the authors
(V.Ch.Zh.) gratefully acknowledges the hospitality of
Prof. M.~Mueller-Preussker and his colleagues at the particle theory
group of the Humboldt University extended to him during his stay
there. We thank also I. E. Frolov for his consultations in some
questions of numerical calculations 
and K. G. Klimenko 
for fruitful discussions. One of us (D.E.) is also grateful to
L.Alvarez-Gaume for useful discussions and comments.
This work was supported by DAAD.


\begin{thebibliography}{999}
\bibitem{rapp}
R. Rapp, T. Sch\"afer, E.A. Shuryak and M. Velkovsky, Phys. Rev.
Lett.
{\bf 81}, 53 (1998); Ann. Phys. {\bf 280}, 35 (2000);
M. Alford, K. Rajagopal and F. Wilczek, Phys. Lett. {\bf B 422}, 247
(1998).
 
\bibitem{NJL}
Y. Nambu and G. Jona-Lasinio, Phys. Rev. \textbf{122}, 345 (1961);
\textbf{124}, 246 (1961); ibid. {\bf 124}, 246 (1961);  V. G. Vaks and A. I. Larkin, ZhETF {\bf
  40}, 282 (1961).
 
\bibitem{MESONS}
D. Ebert and M.K. Volkov, Yad. Fiz. \textbf{36}, 1265 (1982); Z.
Phys. \textbf{C 16}, 205 (1983); D. Ebert and H. Reinhardt, Nucl.
Phys. \textbf{B 271}, 188 (1986).
 
\bibitem{generalN} D. Ebert, H. Reinhardt and M.K.
Volkov, Progr. Part. Nucl. Phys. \textbf{33}, 1 (1994).
 
\bibitem{kunihiro}
T. Hatsuda and T. Kunihiro, Phys. Rep. {\bf 247}, 221 (1994).
 
\bibitem{Ebert_Kaschluhn}
D. Ebert, L. Kaschluhn and G. Kastelewicz, Phys. Lett. \textbf{B
264}, 420 (1991).
 
\bibitem{Vogl}
U. Vogl, Z. f. Phys. \textbf{A 337}, 191 (1990);  U. Vogl and W. Weise,
Progr. Part. Nucl. Phys. \textbf{27}, 195 (1991).
 
\bibitem{Kawati}
S. Kawati and Miyata, Phys. Rev. \textbf{D 23}, 3010 (1981); V.
Bernard, U.-G. Meissner and 
I. Zahed, Phys. Rev. \textbf{D 36}, 819
(1987); Chr.V. Cristov and K. Goeke, Acta Phys. Pol. \textbf{B 22},
187 (1991); D. Ebert, Yu.L. Kalinovsky, L. M\"unchow and M.K. Volkov,
Int. J. Mod. Phys. \textbf{A 8}, 1295 (1993); K.G. Klimenko and
A.S. Vshivtsev, JETP Lett. \textbf{64}, 338 (1996); A.S. Vshivtsev,
V.Ch. Zhukovsky and K.G. Klimenko, JETP \textbf{84}, 1047 (1997).
 
\bibitem{kl}
S.P. Klevansky and R.H. Lemmer, Phys. Rev. {\bf D 39}, 3478 (1989);
H. Suganuma and T. Tatsumi, Ann.of Phys.{\bf 208}, 470 (1991);
T. Inagaki, S.D. Odintsov and Yu.I. Shil'nov,
Int. J. Mod. Phys. {\bf A 14}, 481 (1999);
E. Gorbar, Phys. Lett. {\bf B 491}, 305 (2000).
 
\bibitem{vvk}
M.A. Vdovichenko, A.S. Vshivtsev and K.G. Klimenko,
Phys. Atom. Nucl. {\bf 63}, 470 (2000).
\bibitem{ekvv}
D. Ebert, K.G. Klimenko, M.A. Vdovichenko and A.S. Vshivtsev,
Phys. Rev. {\bf D 61}, 025005 (2000).
\bibitem{ek}
D. Ebert and K. G. Klimenko, Nucl. Phys. {\bf A 728}, 203 (2003).
 
\bibitem{Ebert_Zhukovsky}
D. Ebert, V.V. Khudyakov, V.Ch. Zhukovsky, and K.G. Klimenko,
Phys. Rev. {\bf D 65}, 054024 (2002);
D. Ebert, K.G. Klimenko, H. Toki, and V.Ch. Zhukovsky,
Prog. Theor. Phys. {\bf 106}, 835 (2001).
 
\bibitem{quark_matter}
J. H\"ufner, S.P. Klevansky, P. Zhuang, and H. Voss,
Ann. Phys. (N.Y.) \textbf{234},
225 (1994); I.N. Mishustin et al., Phys. Rev. \textbf{C 62}, 034901
(2000); hep-ph/0010223; M. Hanauske, L.M. Satarov, I.N. Mishustin et
al., Phys. Rev. \textbf{D 64}, 043005 (2001).
 
\bibitem{Ba}
B.C. Barrois, Nucl. Phys. {\bf B 129}, 390 (1977).
 
\bibitem{Frau}
S.C. Frautschi, ``Asymptotic freedom and color superconductivity in
dense quark matter'', in Proceedings of the Workshop on Hadronic
Matter at Extreme Energy Density, Ed., N. Cabibbo, Erice, Italy (1978).
 
\bibitem{bl}
D. Bailin and A. Love, Phys. Rept. {\bf 107}, 325 (1984).
 
\bibitem{one_gluon}
D.T. Son, Phys. Rev.\textbf{ D 59}, 094019 (1999); D.K. Hong, Nucl.
Phys. \textbf{B 582}, 451 (2000); S.D.H. Hsu and M. Schwetz, Nucl.
Phys. \textbf{B 572}, 211 (2000); R.D. Pisarsky and D.H. Rischke,
Phys. Rev. \textbf{D 61}, 074017 (2000); I.A. Shovkovy and L.C.R
Wijewardhana, Phys. Lett \textbf{B 470}, 189 (1999).
 
\bibitem{extream}
K. Rajagopal and E. Shuster, Phys. Rev. \textbf{D 62}, 085007
(2000).

\bibitem{26}
V. A. Miransky, I. A. Shovkovy and L. C. R. Wijewardhana,
Phys. Rev. \textbf {D 62}, 085025 (2000) 
 
\bibitem{Alford}
M. Alford, K. Rajagopal and F. Wilczek, Nucl. Phys. \textbf{B 537},
443 (1999); K. Langfeld and M. Rho, Nucl. Phys. \textbf{A 660}, 475
(1999).
 
\bibitem{Berges}
J. Berges and K. Rajagopal, Nucl. Phys. \textbf{B 538}, 215 (1999).

\bibitem{Schwarz}
T.M. Schwarz, S.P. Klevansky and G. Papp, Phys. Rev. \textbf{C 60},
055205 (1999).
 
\bibitem{Alford_Kebrikov}
M. Alford, hep-th/0102047; B.O. Kerbikov, hep-th/01101975.

\bibitem{shovk}
I.A. Shovkovy, nucl-th/0410091. 
 
\bibitem{braghin}F.L. Braghin, hep-ph/0611390.
 
\bibitem{yud}
D. Ebert, K.G. Klimenko, and V.L. Yudichev, Phys. Rev. {\bf D 72},
056007 (2005); Phys. Rev. {\bf C 72}, 015201 (2005).  
 
\bibitem{blasch}
D. Blaschke, D. Ebert, K. G. Klimenko,M. K. Volkov and V. L. Yudichev,
Phys. Rev. {\bf D 70}, 014006 (2004).
\bibitem{MUTA}
T. Inagaki, S.D. Odintsov and T. Muta, Prog. Theor. Phys. Suppl.
\textbf{127}, 93 (1997), hep-th/9711084 
(see also further references in this review paper). 
 
\bibitem{Elizalde}
E. Elizalde, S. Leseduarte and S.D. Odintsov, Phys. Rev. \textbf{D
49}, 5551 (1994); Mod. Phys. Lett. \textbf{A 9}, 913 (1994).

\bibitem{Elizalde_Shilnov}
E. Elizalde, S. Leseduarte, S.D. Odintsov and Y.I. Shilnov, Phys.
Rev. \textbf{D 53}, 1917 (1996).
 
\bibitem{Gorbar}
E.V. Gorbar, Phys. Rev. \textbf{D 61}, 024013 (1999).
 
\bibitem{Inagaki_Ishikawa}
T. Inagaki and K. Ishikawa, Phys. Rev. \textbf{D 56}, 5097 (1997).
 
\bibitem{ohsaku2}
T. Ohsaku, Phys. Lett. {\bf B599}, 102 (2004).
 
\bibitem{qqrindler}
D. Ebert and V.Ch. Zhukovsky, 
Phys. Lett. \textbf{B645},  267, (2007)
(The paper contains a misprint in the definition of $\gamma$ matrices in
the charge conjugation operator \textit{C}, which should read like this
$\textit{C}=i\gamma^{\hat   2}\gamma^{\hat 0}$. The results of the
paper, however, were obtained with the above correct expression for this
operator and do not depend on this misprint.); 
hep-th/0612009. 

\bibitem{Bunch_Parker}
T.S. Bunch and L. Parker, Phys. Rev. \textbf{D 20}, 2499 (1979).
 
\bibitem{Parker_Toms}
L. Parker and D.J. Toms, Phys. Rev.\textbf{ D 29}, 1584 (1984).

\bibitem{kim_klim}
D.K. Kim and K.G. Klimenko, J.  Phys. \textbf{A 31}, 5565 (1998).
 
\bibitem{Goyal_Dahiya}
A. Goyal and M. Dahiya, J. Phys. G: Nucl. Part. Phys. \textbf{27},
1827 (2001), hep-ph/0011342.
 
\bibitem{Huang_Hao_Zhuang}
X. Huang, X. Hao and P. Zhuang, hep-ph/0602186.
 
\bibitem{brill}
D. R. Brill and J. A. Wheeler, Rev. Mod. Phys. {\bf 29}, 465 (1957).

\bibitem{bordag}
J. S. Dowker, J. S. Apps, K. Kirsten and M. Bordag,
Class. Quant. Grav., {\bf 13}, 2911-2920 (1996); hep-th/9511060. 
 
\bibitem{Camporesi}
R. Camporesi, Phys. Rept. \textbf{196}, 1 (1990); R. Camporesi and
A. Higuchi, gr-qc/9505009.
 
\bibitem{Weinberg}
P. Candelas and S. Weinberg, Nucl. Phys. B 237, 397 (1984).
 
\end{thebibliography}
\end{document}